\documentclass{llncs}

\usepackage[pdftex]{graphicx}
\usepackage{amssymb}
\usepackage[plain]{algorithm}
\usepackage{algorithmic}
\usepackage{cite}
\usepackage[belowskip=-5pt,aboveskip=5pt]{caption}

\begin{document}

\title{Epidemic Information Diffusion: A Simple Solution to Support Community-based Recommendations in P2P Overlays}
\author{Patrizio Dazzi\inst{1} 
    \and Matteo Mordacchini\inst{2} 
    \and Laura Ricci\inst{3}
}
\institute{ISTI-CNR, Pisa, Italy \\ \email{patrizio.dazzi@isti.cnr.it}
\and IIT-CNR, Pisa, Italy \\ \email{matteo.mordacchini@iit.cnr.it}
\and University of Pisa \\ \email{matteo.mordacchini@iit.cnr.it}
}

\maketitle

\begin{abstract}
Epidemic protocols proved to be very efficient solutions for supporting dynamic and complex information diffusion in highly distributed computing infrastructures, like P2P environments. They are useful bricks for building and maintaining virtual network topologies, in the form of overlay networks as well as to support pervasive diffusion of  information when it is injected into the network. This paper proposes a simple architecture exploiting the features of epidemic approaches to foster a collaborative percolation of information between computing nodes belonging to the network aimed at building a system that groups similar users and spread useful information among them.
\end{abstract}

\section{Introduction}
This paper proposes the recipe for the definition of a simplified system architecture that aims at exploiting a collaborative exchange of information between peers belonging to a highly distributed infrastructure in order to build a computing/network approach able to link similar users in order to foster the process of information percolation within the nodes of the network.
More in detail, we push further the idea of realising collaborative recommender mechanisms, by means of solutions fostering interest clustering, that are obtained by means of interactions happening among users.
Our approach couples epidemic-based P2P overlay networks to ease the gathering of users with similar interests and then use the connections established to let peer exchange recommendations one each others. 
Our goal is twofold. On one hand we aim at building an adaptive system supporting the recognition of communities of users\' interests in a decentralized, distributed way. The approaches that have been proposed so far in the area of P2P computing (and the epidemic ones, in particular) are able to manage a very large amount of peers and to deal gracefully with churn, whereas centralized systems require expensive and, often, very complex techniques to ensure continuous operation under node and link failures. 
The service is implemented by means of the collaborations established between computing nodes, without needing any centralized authority devoted to store all the profiles and the ratings of users as well as to provide centralized-controlled suggestions.
On the other hand, our goal is to exploit such communities not limiting our aim to the knowledge sharing about interesting items within them, but also to address some of the traditional problems affecting recommender systems. In particular, the ability to recommend new, almost unknown, items. 
The system we are sketching, assumes that each neighbor of a computing node (peer) $P$ pushes recommendations to it focused on the items that might be of potential interest for $P$. It is worth to notice that this decision is taken locally, when a neighbor selects or becomes aware from its links of the existence of a new item, whose characteristics, are someway related with one (or more) of its communities. 
Then, it can then recommend such item both to $P$ and to its other neighbors, of all the related communities, as well. This approach would allow a more efficient and rapid percolation of the information within the overlay network.
The remain of this paper is organized as follows: in Sec.~\ref{rel} we shortly present the scientific literature about the subject of this paper; in Sec.~\ref{arch} we describe the architecture of our proposed system. Finally, in Sec.~\ref{concl}, conclusions are given and potential further exploitations of this work are proposed.

\section{Related Work}\label{rel}

%\cite{}

The correlations of interests amongst a group of distributed users has been leveraged in a variety of contexts and for designing or enhancing various distributed systems~\cite{carlini2012reducing, carlini2011evaluating}. 
For peer-to-peer file sharing systems that include file search facilities (e.g., Gnutella, eMule, etc.), an approach to increase recall and precision of the search is to group users based on their past search history or based on their current cache content ~\cite{Fraigniaud2005Combining-the-use-of,Handurukande2006Peer-Sharing-Behavio}. 
Another potential use of interest clustering is to form groups of peers that are likely to be interested in the same content in the future, hence forming groups of subscribers in a content-based information diffusion system~\cite{carlini2009service,baraglia2013peer,dazzi2009peer,baraglia2011group,mordacchini2009challenges,dazzi2011experiences,carlini2014toward,mordacchini2013towards,marzolla2006p2p,gennaro2007mroute}.
Moreover, interest correlation can be used to help bootstrapping and self-organization of dissemination structures such as network-delay-aware trees for RSS dissemination~\cite{Patel2009Rappel:-Exploiting-i}. The correlation between the users\' past and present accesses has been used for user-centric ranking. 
In order to improve the customisation of search results, the most probable expectations of users are determined using their search log stored on a centralized server~\cite{tan_mining_2006,teevan_personalizing_2005}. However, the correlation between users is not leveraged to improve the quality of result personalization, hence making the approach viable only for users with sufficiently long search logs. An alternative class of clustering search engines uses semantic information in order to cluster results according to the general domain they belong in (and not as in our approach to cluster users based on their interests). This can be seen as a centralized, user-agnostic approach to improve user experience. The clustering amongst data elements is derived from their vocabulary. It presents the user with results along different interest domains and can help the user to disambiguate these results from a query that may cover several domains, e.g., the query word apple can relate to both food/fruits and computers domains. Examples of such systems are EigenCluster~\cite{Cheng2006A-divide-and-merge-m}, or TermRank~\cite{Gelgi2007Term-Ranking-for-Clu}. Nonetheless, these systems simply modify the presentation of results so that the user decides herself in which domain the interesting results may fall these results are not in any way automatically tailored to her expectations. They do not also consider the clustering of 
interest amongst users, but only the clustering in content amongst the data.

Other approaches cluster users on the basis of similarity between their semantics profile. Approaches of this kind of systems includes GridVine~\cite{GridVine}, the semantic overlay networks~\cite{SON} and p2pDating~\cite{p2pDating}. 
They build a semantic P2P overlay infrastructure that relies on a logical layer storing data. 

They make use of heterogeneous but semantically related information sources whereas our approach does not rely on any kind of semantic interpretation. It, in principle, enables a broader exploitation of more heterogeneous data sources. 
Related with our proposal is Tribler~\cite{tribler}, a P2P television recommender system. In contrast with our approach, neighbor lists can be directly filled in by the user herself using an interface. No topology or affinity property is considered. We propose a gossip system that construct and maintain in rest groups of dynamic users based on their past activities, without needing their direct intervention.

\section{Proposed approach}\label{arch}
\begin{figure}[ht!]
\begin{minipage}[t]{20em}
\begin{center}
\includegraphics[width=0.5\columnwidth]{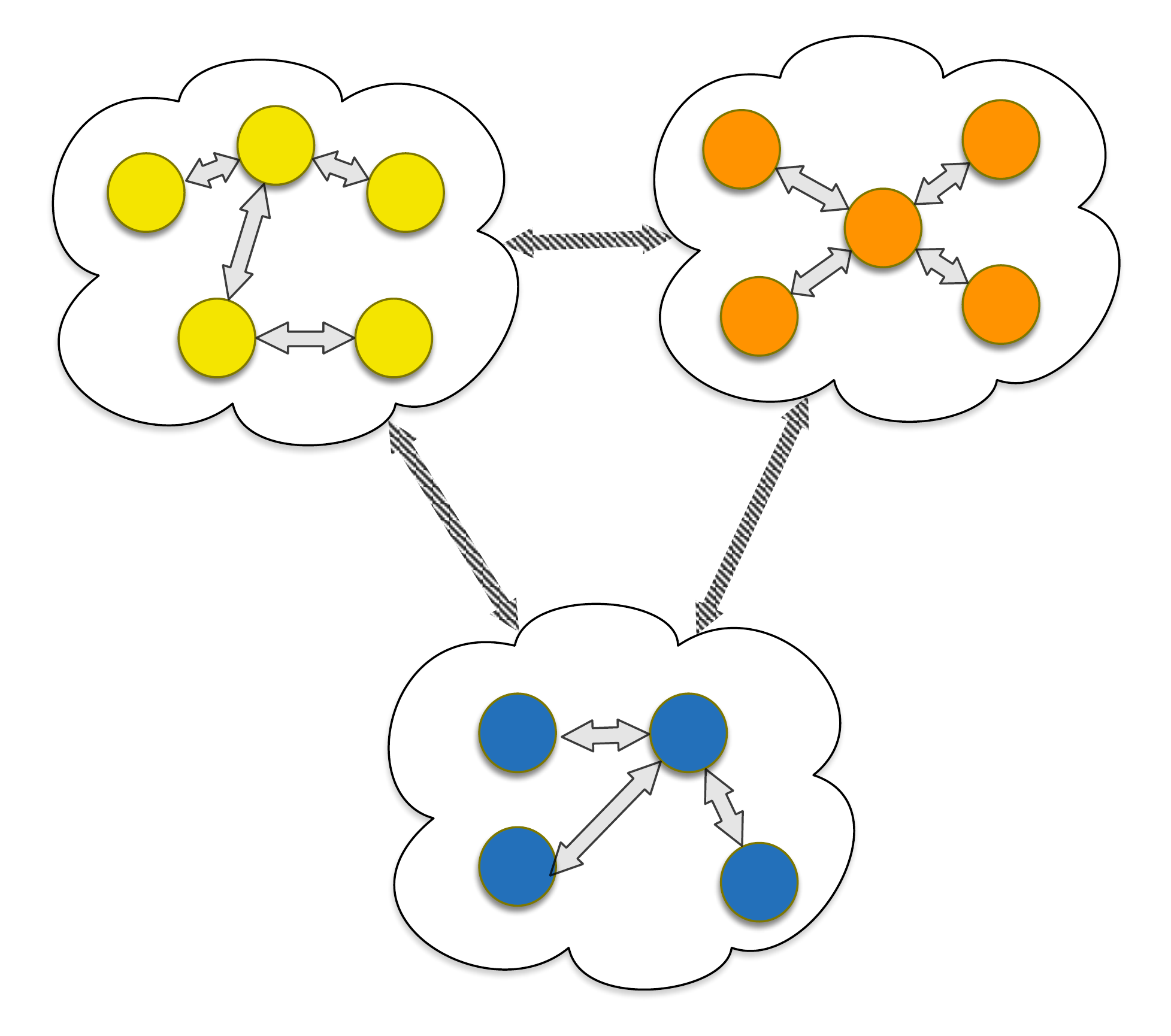}
\caption[Interest Communities]{Interest Communities}\label{fig:communities}
\end{center}
\end{minipage}
\hfill
\begin{minipage}[t]{20em}
\begin{center}
\includegraphics[width=0.5\columnwidth]{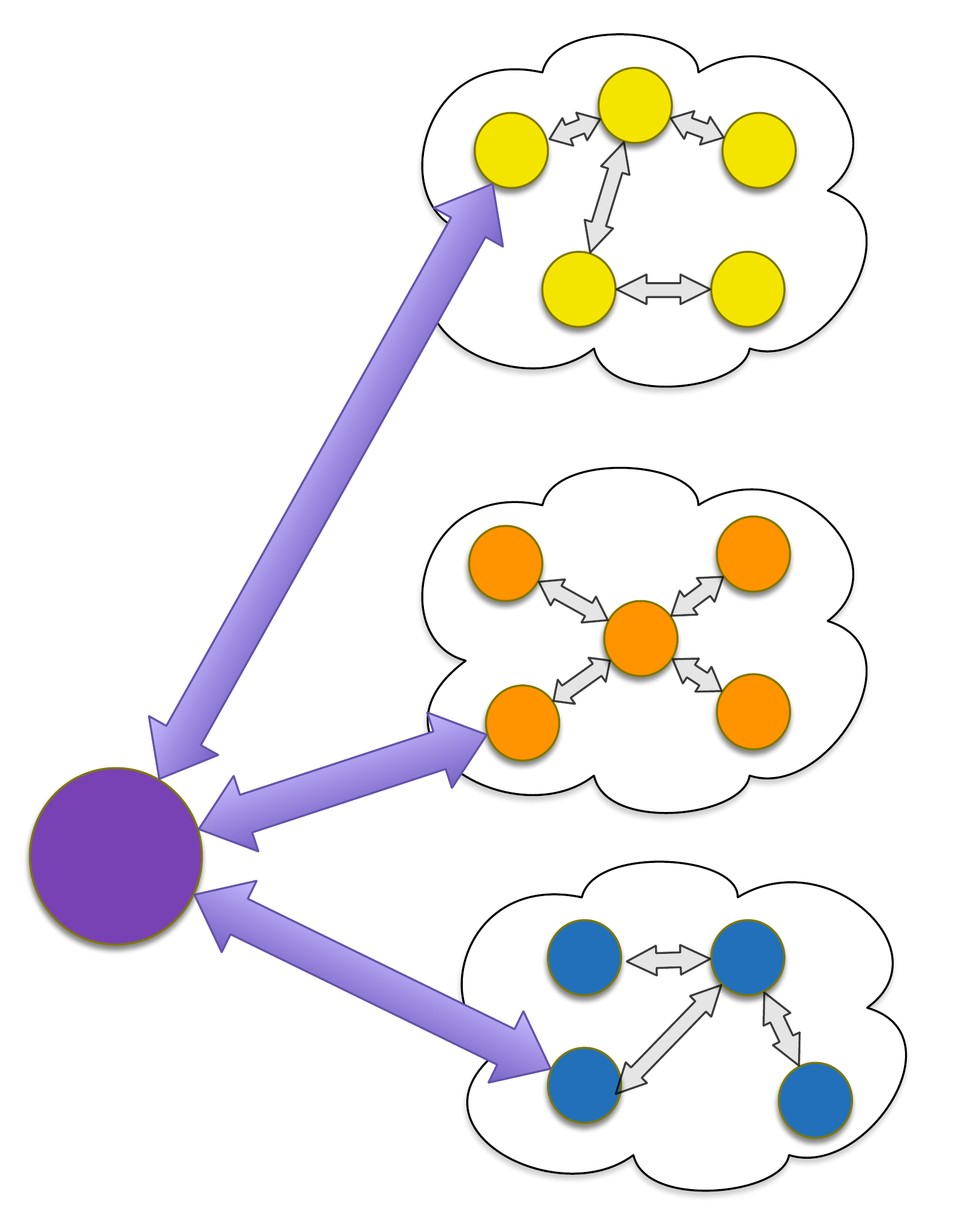}
\caption[Interest Overlays]{Interest Overlays}\label{fig:interests}
\end{center}
\end{minipage}
\end{figure}

This section introduces the main pillars that would be needed to support the construction and the exploitation of an overlay network made of peers that share common interests. 
Figure~\ref{fig:communities} sketches the architecture of an overlay network organised accordingly. 
As can be observed in figure, the links between peers are established when they are characterised by a common interest. This information is derived recognising the accesses performed by peers to the same content in the past. 
The surrounding idea is that they are considered interested to share similar interest if can potentially show interests for the same content in the future. Thus, peers collaboratively exchange useful recommendations among themselves.

The protocol, to group similar users in communities, adopts a clustering algorithm. As a first step each peer determines, independently, the peers to link with. 
These one-to-one connections are established on the basis of an interest-based degree, that is measured amongst the peer it encounters. Every time it becomes aware of a new peer, it can, in turn, learn of the existence of new potential neighbors and possibly communicate with them. Finally, it can also be aware of other, potentially better neighbors. 
The idea is that this process is stabilized when each node composing the neighborhood of a peer can be considered as the representative of a community of a shared interest. An important side-effect of this vision is that a peer is characterized by multiple interests, with different ``entry-points'' for each interest. 
The process is conducted separately for each of the interests of a peer. Consequently, the connections are established and maintained separately for every distinct interest. At the end will be created a set of virtual different overlays, where each peer participates in as many groups is required to cover its interests.
The resulting scenario situation is depicted in Fig.\ref{fig:interests}. In order to obtain such organisation, each peer initiate a different stream of messages, one for each of its interests.

\subsection{Profiles}
Profiles of peers need to be modeled according to the users' interests. A possible approach would be based on recently accessed resources, purchased items, visited pages, etc.
Such information, once gathered, has to be considered in a proper way. Basically, it consists in the basis over which the overlay network will be organised. 
Generally speaking, let $\Im$ be the set of items belongings to the whole set of profiles of users and let $\Im_p \subseteq \Im$ be the subset of items belonging to a specific peer $p$. We consider that the profile $\pi$ of $p$ can be defined as
\begin{eqnarray*}
\pi_p = \{ (i, C(i), R(i)) | i \in \Im_p \}
\end{eqnarray*}
where $i$ is an item belongings to the set of $\Im_p$, $C(i)$ is the content associated with $i$ and $R(i)$ is the rating given by $p$ to $i$. The peer $p$ has also associated a set $I^p = \{I^{p}_1,\ldots,I^{p}_k\}$ of interests. 
Each of the items in $\pi_p$ may be associated with an interest $I^{p}_j$. We can then represent $\pi_p$ in the following way:
\begin{eqnarray*}
\pi_p = \displaystyle\bigcup_{j=1,...,k} \pi_{p}(I^{p}_j)
\end{eqnarray*}
where $\pi_{p}(I^{p}_j) $ is the set of items related to the interest $I^{p}_j$. 
For realising this association, we introduce a function $\gamma$ that given an item belonging to $\Im$ decides the interest it should be associated with. More formally:
\begin{eqnarray*}\gamma_p(i) = I^p_j~~with~i \in \Im_p\end{eqnarray*}
It is worth to note that the set $I^p$ is specific for each distinct peer $p$. 
In fact, we do not assume any globally known labeling, classification or partitioning of the objects in $\Im$. Each peer performs its own subdivision of $\Im_p$ in the interests of $I_p$. 
It can then compare its objects divided per interest with the sets of the other peers it will contact. Given two peers $p_1$ and $p_2$, $p_1$ would consider its local interest $I^{p_1}_s$ similar to the interest $I^{p_2}_t$ if it would contain the most similar set of items among the other sets in $I^{p_2}$ with respect to the items in $I^{p_1}_s$.
As a consequence of having a solution to describe each user interests coded in the peer profiles, it is important to pay attention on adopting a proper similarity function $sim: \Pi^2 \rightarrow \mathbb{R}$ to compare profiles, where $\Pi$ is the set of all possible profiles. 
This is a key aspect, since this function specify the relationships between peers according to their interests. 
If each distinct interest is determined by different type of features, different similarity measures could be used to evaluate peers proximities with respect to each interest.
Several measures can be adopted to this end. As an example, a typical approach is to use a metric that takes into account the size of each profile, such as the Jaccard similarity, which has proven to be an effective similarity measure~\cite {Fraigniaud2005Combining-the-use-of,Patel2009Rappel:-Exploiting-i}. Given two peers $p_1$ and $p_2$ and two interests $I^{p_1}_s$ and $I^{p_2}_t$, the similarity  can be computed as 
\begin{eqnarray*}
sim(p_1,p_2) = \frac{|\pi_{p_1}(I^{p_1}_s) \cap \pi_{p_2}(I^{p_2}_t)|}{|\pi_{p_1} (I^{p_1}_s)\cup \pi_{p_2}(I^{p_2}_t)|}
\end{eqnarray*}

\subsection{Setup of Interest Communities}
One of the base assumptions of our envisioned system is that every peer is able to compute its interest-based distance to any other peer in the network. This measure allows it drive is ability to \emph{group} with other peers that have close-by interests, in order to form the basis for \emph{interests communities}. 
This process is conducted automagically in a self-organizing and completely decentralized way, using a epidemic communication. 
Each peer knows a set of other peers, namely its neighbors, and tries periodically to choose new neighbors that are closer to its interest than the previous ones. 
In our envisioned system, this is simply obtained by discovering new peers from some other peer, then retrieving their profiles. Finally, choosing the $C$ nearest neighbors in the union of present and potential neighbors. 
When a peer $p$ joins the network, it becomes in contact with one or more peers already belonging to in the interest-proximity network overlay. 
They use the profile similarity function to compute how similar they are. They consider each interest in the $I$ of $\pi_p$ separately and they compare it against their own. 
Furthermore, the peers contacted by $p$ use the same similarity function to determine which are, among their neighbors, the most similar to $p$. Once determined, the join request of $p$ are routed toward them. 
All the peers that receive that request will react using the same protocol described above. All the interactions are shown in Algorithms 1 and 2. 
This approach will lead $p$ to become aware of the existence of the most similar peers in the network overlay and allow it to connect with them. In doing this process, the involved peers can only use their local knowledge to compare their respective profiles.
\begin{table}[tb] 
\begin{tabular}{l|l} 
\begin{minipage}{2.35in}
{\scriptsize\textbf{Algorithm 1}}
\begin{algorithmic}
{\scriptsize
\STATE Let $CR(P')$ be a connection request from another peer $P'$
\STATE Let $NewPeers = \emptyset$
\IF{$Sim(P,P^{'}) \geq \displaystyle\min_{P_i \in N(P)}Sim(P,P_i)$}
\STATE Accept $CR(P')$
\FORALL{$P_i \in N(P)$}
\IF{$Sim(P_i,P') \geq \theta$}
%\STATE send to $P^{'}$ its most sim. peers in $N(P)$
\STATE add $P_i$ to $NewPeers$
\ENDIF
\ENDFOR
\STATE add $P'$ to $N(P)$
\STATE send $NewPeers$ to $P'$
\ELSE
\STATE refuse $CR(P')$
\ENDIF
}

\end{algorithmic}\label{alg:1}
\end{minipage} 
& 
\begin{minipage}{2.35in}
{\scriptsize\textbf{Algorithm 2} }
\begin{algorithmic}                    % enter the algorithmic environment
{\scriptsize
\STATE Let $N(P)$ be the set of $P'$s actual neighbors
\FORALL{$P_i \in N(P)$}
\STATE Get from $P_i$ a set $NewPeers$ from its neighborhood
\FORALL{$P^{'} \in NewPeers$}
\IF{$P^{'} \notin N(P)$}
\STATE connect with $P^{'}$
\IF{$Sim(P,P^{'}) \geq \displaystyle\min_{P_j \in N(P)}Sim(P,P_j)$}
\STATE add $P^{'}$ to $N(P)$
\ENDIF
\ENDIF
\ENDFOR
\ENDFOR 
}
\end{algorithmic} 
\end{minipage} \label{alg:2}

\\ 
\\ \hline
\\

\begin{minipage}{2.35in} 
{\scriptsize\textbf{Algorithm 3}}
\begin{algorithmic}
{\scriptsize
\STATE Let $N_p(I_j)$ be the set of $P'$s neighbors for the interest $I_j$
\STATE Receive a recommendation request from $p' \in N_P(I_j)$
\FORALL{$i \in \pi_p(I_j)$}
\IF{$Sim(p',i) \geq \theta$}
\STATE recommend $i$ to $p'$
\ENDIF
\ENDFOR
}
\end{algorithmic} \label{alg:3}
\end{minipage} 
&
\begin{minipage}{2.35in} 
{\scriptsize\textbf{Algorithm 4}}
\begin{algorithmic} 
{\scriptsize
\STATE Know about a new item $h$
\STATE Let $I_j$ be the interest $h$ is related to
\STATE Let $N_P(I_j)$ be the neighborhood of peers interested in $I_j$
\FORALL{$p' \in N_P(I_j)$}
\IF{$Sim(p',h) \geq \theta$}
\STATE recommend $h$ to $p'$
\ENDIF
\ENDFOR
}
\end{algorithmic} \label{alg:4}
\end{minipage} 
\end{tabular}
\\ \\
\caption{Active and passive threads and pull and push recommender algorithms}
\vspace{-15pt}
\end{table} 

Once the process is stabilized, $p$ can consider its neighbors as the representatives of a personal community of ``friends'' from which request and to which forward recommendations. Thus, the gossip protocol provide the basis for classical recommender systems in forming the set of similar users. 
This is done distributively and adaptively and the epidemic protocol ensure a robust and constant maintenance over time. Recommendations can then be requested by $p$ to its neighborhood and it can forward the newly items it discovered to its neighbors using Algorithms 3 and 4. 

\section{Conclusion}\label{concl}
The focus of this paper is on giving a simple recipe for addressing the problem of clustering users in a purely decentralized way to foster information exchange. This is a particularly useful brick for enabling self-emerging and automated creation of communities of nodes representing users, which share common interests. In this paper we sketched the overall architecture of a epidemic-based distributed system exploiting a collaboratively built recommender system. The solution sketched in this work is simpler than most part of the existing solutions. This is inline with our goal: keep the solution as simple as possible but still providing a solution that exploit collaborative filtering is able to provide recommendations that are tailored and offer an acceptable degree of serendipity.

\bibliographystyle{unsrt}
\bibliography{newbiblio}

\end{document}